# Detecting Permanent and Intermittent Purchase Hotspots via Computational Stigmergy

Antonio L. Alfeo[1], Mario G. C. A. Cimino[1], Bruno Lepri[2],
Alex "Sandy" Pentland[3], and Gigliola Vaglini[1]

[1] *Department of Information Engineering, University of Pisa, Largo Lazzarino 1, Pisa, Italy*
[2] *Bruno Kessler Foundation, via S. Croce, 77, Trento, Italy*
[3] *M.I.T. Media Laboratory, 75 Amherst Street, Cambridge 02142, USA*
{luca.alfeo,mario.cimino,gigliola.vaglini}@ing.unipi.it,lepri@fbk.eu,pentland@mit.edu

Abstract: The analysis of credit card transactions allows gaining new insights into the spending occurrences and mobility behavior of large numbers of individuals at an unprecedented scale. However, unfolding such spatiotemporal patterns at a community level implies a non-trivial system modeling and parametrization, as well as, a proper representation of the temporal dynamic. In this work we address both those issues by means of a novel computational technique, i.e. computational stigmergy. By using computational stigmergy each sample position is associated with a digital pheromone deposit, which aggregates with other deposits according to their spatiotemporal proximity. By processing transactions data with computational stigmergy, it is possible to identify high-density areas (hotspots) occurring in different time and days, as well as, analyze their consistency over time. Indeed, a hotspot can be permanent, i.e. present throughout the period of observation, or intermittent, i.e. present only in certain time and days due to community level occurrences (e.g. nightlife). Such difference is not only spatial (where the hotspot occurs) and temporal (when the hotspot occurs) but affects also which people visit the hotspot. The proposed approach is tested on a real-world dataset containing the credit card transaction of 60k users between 2014 and 2015.

## 1 INTRODUCTION

The extensive usage of nowadays pervasive technologies generates a large number of digital traces associated with each human activity. Few well-known examples are social media posts (Cimino, Lazzeri, Pedrycz, & Vaglini, 2018), vehicle GPS traces (Alfeo et al., 2018), mobile phone records (Louail, et al., 2014), smart cards usage (Zhong, Manley, Arisona, Batty, & Schmitt, 2015), and credit card transactions (Dong, Meyer, Shmueli, Bozkaya, & Pentland, 2018). Among the many possible sources, transactions' datasets can provide insights about the daily activities of large numbers of individuals, allowing the analysis of both their spending occurrence and mobility behavior at unprecedented scale (Dong X. et al., 2018). Indeed, individuals' purchases result from the combination of their needs, habits, well-being and where they can go shopping. Moreover, each shopping choice drives individuals' movement (Krumme, Llorente, Cebrian, & Moro, 2013). By analyzing both the spending occurrences and the mobility patterns of the consumers it is possible to gain new insights about individuals' behavior (Singh, Bozkaya, & Pentland, 2015), as well as understanding the structure and usage of a given urban area (Long & Liu, 2016). One method for such urban areas characterization employs the detection of purchase hotspots, i.e. locations with a significant occurrence of purchase events (Sobolevsky, et al., 2015). By means of a hotspots analysis, many works in the field address the managing of different urban operational problem (Sobolevsky, et al., 2014), such as transportation services' demand (Fuchs, Stange, Hecker, Andrienko, & Andrienko, 2015) (Oh, Cheng, Lehto, & O'Leary, 2004). However, the results provided by the hotspots' detection should also take into account the hotspot dynamics, i.e. their changing over space and time (Brimicombe, 2005). Indeed, hotspots' occurrence may be due to large time-scales (e.g. seasonal, weekly) regularities, or according to real world events such as holidays and sales (Uncles, Ehrenberg, & Hammond, 1995). In this context, there has not been sufficient research on characterizing hotspots from a dynamic perspective (Khan, Bergmann, Jurdak, Kusy, & Cameron, 2017). To address this problem, we propose an approach aimed at unfolding purchase hotspots and characterize their spatial and temporal dynamics. Our approach employs a novel computational technique based on the principle of Stigmergy, a self-organization mechanism used by social insect colonies and based on the deposit of pheromone marks. Since pheromones are volatile, a pheromone trail (i.e. the marks aggregation) appears only in areas characterized by a consistent deposit activity. By applying this pheromone-like aggregation to purchase event occurrences, the resulting trail is able to summarize their spatiotemporal density, enabling the detection of hotspots. Such approach is known as *Computational Stigmergy*, and can be used to detect the hotspots (Alfeo, Cimino, Egidi, Lepri, & Vaglini, 2018) and characterize their occurrence over time according to their similarity in different time instants. As an example, this similarity between hotspots can be extracted from data, and then used to carry out a clustering process on the corresponding relational space (Cimino, et al., 2006). We test the proposed approach with a real world dataset provided by a large Turkish financial institution's, consisting of all credit card transactions made in 12 months between 2014 and 2015 from more than 60k customers. In the following sections we present our approach and the results obtained by analyzing those data. In Section 2 we briefly present a literature review about the approaches based on hotspot analysis. In Section 3 we detail the proposed approach, whereas in Section 4 the experimental setup and the obtained results are discussed. Finally, we draw the conclusions of this study in Section 5.

## 2 RELATED WORKS

The notion of hotspots, i.e. a location with relatively high levels of activity, was firstly used in order to understand the occurrence of criminal activities (Sherman, Gartin, & Buerger, 1989) (Chainey, Tompson, & Uhlig, 2008). Nowadays, the usage of the concept of hotspot has been extended to a number of different context such as epidemiology (Martinez-Urtaza, et al., 2018), transportation (Alfeo A., et al., 2017), and social science (Zhu & Newsam, 2016).

Indeed, the hotspot detection and analysis is more and more exploited by researchers, thanks to its ability of summarize and gain insights into complex phenomena, resulting in applications ranging from urban planning (Liu, Wang, Xiao, & Gao, 2012) to behavioral analysis and activities forecasting (Dong, et al., 2018) (Scholz & Lu, 2014). In this context, two main approaches are used to detect the hotspots: those are respectively based on a statistical and density-based characterization the occurrences (e.g. smart card usage, trips) under investigation.

As an example, an approach of the first group aims at detecting hotspots by employing spatial autocorrelation indicators (Yang, Zhao, & Lu, 2016). In (Yu & He, 2017) authors exploit a heat map to study the discrete distribution of travel demand at the bus stops in order to unfold trip hotspots. However, one of the main difficulties with these approaches remains the inclusion of the time domain in the analysis (Klemm, et al., 2016).

The second group of approaches aimed at detecting hotspot employs the concept of spatial and temporal density. As an example, in (Hu, Miller, & Li, 2014) authors exploit a kernel density approach to represent multiple mobile objects as a density surface and extract its geometric features to analyze the hotspot distribution. Again, in (Senaratne, Broring, Schreck, & Lehle, 2014) authors utilizes a kernel density estimation to detect hotspot clusters of social network activities and analyze their trajectory over time, allowing the detection of urban events (e.g. concerts). However, two shortcomings characterize most of these approaches: (i) the temporal component is neglected or represented by just adding a further dimension to the problem, thus a proper representation and analysis of the temporal dynamics is missing (Yuan, Sun, Zhao, Li, & Wang, 2017); (ii) the hard-to-manage exploration of different analysis parametrization, which is a fundamental feature since each phenomenon is visible at a specific scale and resolution (Atluri, Karpatne, & Kumar, 2018) (Yuan & Raubal, 2012). In this work, we address both those issues by using Computational Stigmergy. Indeed, Computational Stigmergy intrinsically embodies the time domain (Barsocchi, et al., 2015) and allows adapting the analysis by tuning its structural parameters (Alfeo, et al., 2018).

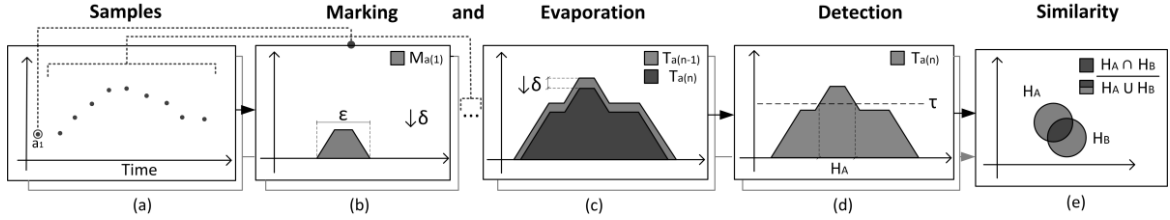

Fig. 1. Architectural modules of the mechanism aimed at detecting hotspots via Computational Stigmergy.

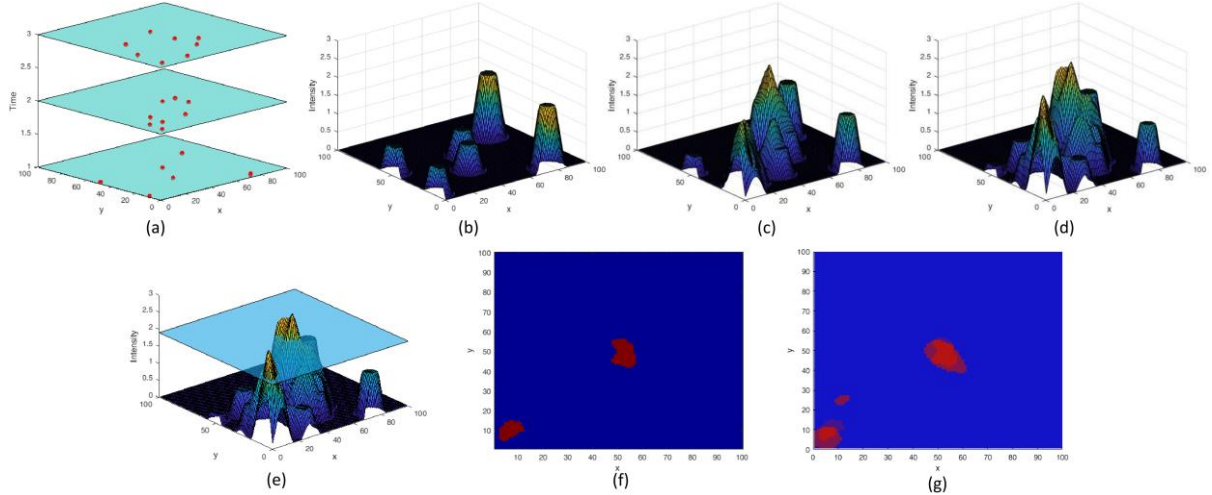

Fig. 2. Samples processing phases. Each sample in a given time instant (a) is transformed into a mark (b); each mark aggregates with the marks released in previous time instants and properly "evaporated" (c-d); a threshold is used to identify the areas corresponding to the significant part of the trail (e-f); those areas are compared to measure their similarity (g).

## 3 FUNCTIONAL DESIGN

The proposed approach is based on the principle of stigmergy, a self-organization mechanism used in ants' colonies (Marsh & Onof, 2008). With stigmergy, the occurrence of a specific condition (e.g. an individual discovering food) corresponds to the release of a pheromone mark in a shared environment. Due to the volatility of the pheromones, isolated marks evaporate and eventually disappear, whereas marks subsequently deposited in proximity to each other aggregate, resulting in a long-lasting and stable pheromone trail. By following the pheromone trail, the colony is steered toward the region in which the condition above (e.g. the discovery of food) occurs consistently, since only such consistency generates the density of marks needed to generate a stable trail.

To the end of unfolding spatiotemporal density in the data, we process them by employing such mechanism. The overall processing schema is known as *Computational Stigmergy* (Alfeo A., et al., 2017). Specifically, with *Computational Stigmergy*, a virtual pheromone deposit (i.e. a mark) is released in a virtual environment (Fig. 1b) in correspondence to the location and time of appearance of each data sample (Fig. 1a), i.e. a credit card transaction. The marks are represented by a truncated cone with a given width $\varepsilon$ and intensity (height). Moreover, marks are subject to an evaporation process, i.e. a temporal decay with rate $\delta$ (Fig. 1c). Thanks to the evaporation, isolated mark progressively disappears, whereas marks that are frequently released in proximity to each other aggregate forming the trail. In a nutshell, the trail appears and stays only in correspondence of consistent marks depositing activity, thus can be considered as a summarization of spatiotemporal density in the data (Alfeo A., et al., 2017). Eq. 1 describes the trail $T_i$ at time instant $i$, resulting by the evaporation of the trail $T_{i-1}$ and the aggregation of the set of *Marks$_i$* at the time instant $i$.

$$T_i = T_{i-1} - \delta + Marks_i \quad (1)$$

As shown in Fig. 1d, in order to detect the significant part of the trail (i.e. corresponding to a potential hotspot $H_i$ at time $i$), it is necessary to set a threshold $\tau$, i.e. a given percentage of the maximum intensity of the trail (Eq. 2).

$$H_i = T_i > \tau \quad (2)$$

Different hotspots can be compared by means of the Jaccard similarity (Niwattanakul, Singthongchai,

Naenudorn, & Wanapu, 2013). Such measure of similarity is computed as the ratio between the intersection (∩) and the union (∪) of the areas underlying the hotspots (e.g. $H_A$ and $H_B$ in Fig. 1e and Eq. 3), and it is defined between 0 (completely different hotspots) and 1 (identical hotspots).

$$S_{H_A,H_B} = \frac{|H_A \cap H_B|}{|H_A \cup H_B|} \quad (3)$$

Thanks to this similarity measure it is possible to identify hotspots on the basis of their temporal consistency, i.e. their similarity in different time instants. For the sake of clarity, we depict the phases of the hotspot detection and analysis with a set of samples belonging to 3 consecutive time slots in Fig. 2. Each of the presented processing steps is parameterized, and each parameter represents a feature of the potential hotspot. Specifically, the marks width Ɛ and the evaporation δ results in the spatial and temporal proximity allowing marks to aggregate and form the trail; by tuning such parameters we define the spatio-temporal density of a potential hotspot. On the other hand, the threshold τ defines the significant amount of spatiotemporal density corresponding to a hotspot. By employing a heuristic or a measure of the quality of the hotspots' detection process, it is possible to tune those parameters with the aim at specialising the detection of the hotspots to the peculiarities of the scenario under analysis, as specified in Section 4.

## 4 EXPERIMENTAL RESULTS

The approach presented in Section 3 has been developed in Matlab, a well-known versatile high-performance environment equipped with functions and syntax to work with big data. We experiment our approach on a credit card transactions dataset provided by a major financial institution in Turkey. Such dataset consists of more than 10 million transactions instances made by more than 64k individuals during a period of twelve months. Each instance is composed by the following attributes: customer id, timestamp, amount, shop id, online, expense type, currency, latitude coordinates, and longitude coordinates. Moreover, we work with an additional dataset containing a number of demographic information about each individual, such as age, gender, education level, income, home, and work location. The customer-level data are anonymized by representing each customer with a pseudo-unique number. By exploring the spatial distribution of the transactions in the dataset it is evident that most of them occur in Istanbul metropolitan area, thus we decide to focus the analysis on such area. The data undergo a pre-processing phase consisting of the spatial and temporal discretization. Each bin corresponds to a square area of 100 meters side and a time interval of 20 minutes. As mentioned in Section 2, the occurrence of hotspots alone provides an incomplete overview of the purchase phenomenon (Uncles, Ehrenberg, & Hammond, 1995). What is really interesting is the consistency of the hotspots' over time. Indeed, a hotspot can be *permanent*, i.e. present throughout the period of observation, or *intermittent*, i.e. present only in certain times and days of the week (Louail, et al., 2014). Most of the works in the field detect a weekly hotspot routine (Alfeo A. , et al., 2018). Thus, we define two types of days (weekends and weekdays) and we split each day into 12 time slots, 2 hours each. In contrast to permanent hotspots, the intermittent hotspots are peculiar for each tuple [*day type; time slot*] within the same month.

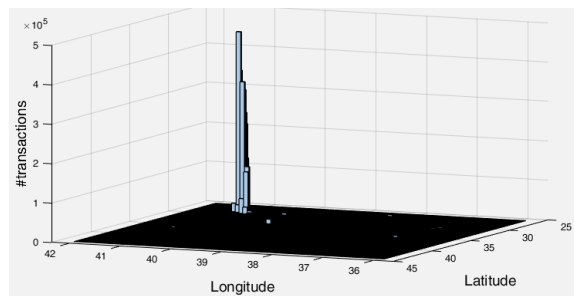

Fig. 3. Transactions' location (GPS) spatial distribution.

To detect both intermittent and permanent hotspots, a number of parameters have to be properly set. The evaporation must guarantee the preservation of (even a part of) the information (i.e. the mark) for the whole time of the analysis. If looking for permanent hotspots, this time corresponds to a whole day, while with intermittent hotspots, it corresponds to 2 hours (a time slot). Thus, in the first case, the evaporation $\delta_P$ is set to 0.01 whereas in the former one $\delta_I$ is set to 0.15. The mark intensity is set to 1, whereas its width Ɛ is set to 10, enabling the aggregation of marks which distance is up to 1 km. The most sensitive parameters of the analysis are the thresholds for permanent $\tau_P$ and intermittent $\tau_I$ hotspots. For this reason, they are set through an iterative exploratory analysis aimed at maximizing only the similarity between intermittent hotspots of similar temporal tuples. As an example, in Fig. 4 we show a similarity matrix obtained from the comparison of the intermittent hotspots for each time slot of each day of September 2014. To a lighter color corresponds a greater similarity (1 on the diagonal). Clearly, there is a consistency of the hotspots (a greater similarity) in correspondence of time windows corresponding to similar occurrences. For example, it is possible to identify a strong similarity between weekend daytimes hours, the evening and night hours between Saturday and Sunday and the daytime hours of the working days. The clearer the distinction between these groups of days, the better the parameterization of the analysis.

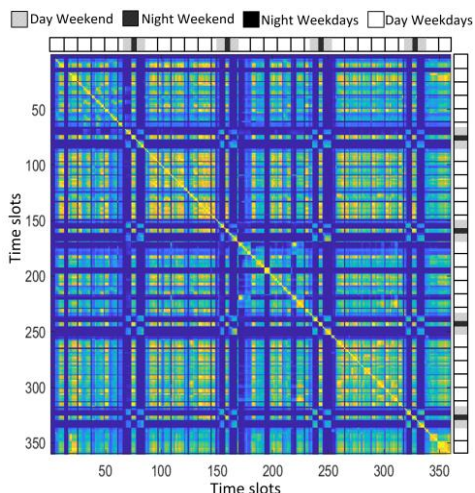

Fig. 4. Similarity matrix obtained by matching the Intermittent hotspots during September 2014.

By definition, intermittent hotspots of a given tuple are supposed to be similar to each other within the same month. On the other hand, the permanent hotspots occur on average all days and time slots, thus their presence may interfere with the similarity computation between hotspots obtained with different tuples. For this reason, we (i) firstly, identify permanent hotspots as the intersection of the areas underlying the significant part of the trails in all days, then (ii) we remove the transactions occurring in those areas, and finally (iii) intermittent hotspots are targeted in order to maximize the similarity between hotspots of the same tuples or temporally close tuples.

Fig. 5 shows the procedure described above together with the 10 permanent (in red) and the 9 intermittent (in cyan) hotspots discovered.

In the absence of a ground truth on the distribution and nature of the hotspots, we discuss the obtained results according to the features of the people spending in the hotspots. Specifically, it is possible to evaluate the characteristics of the users using the permanent or intermittent hotspots from the demographic data associated with the transactions dataset.

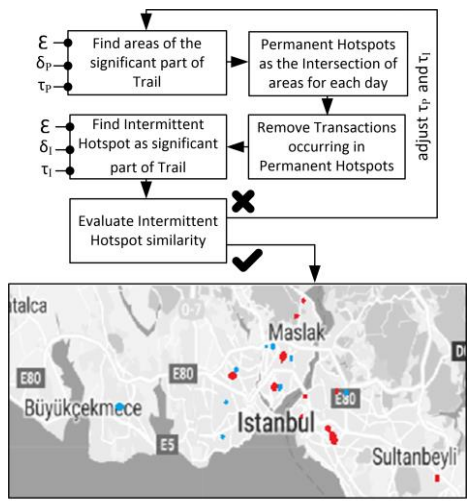

Fig. 5. Discovery process and resulting intermittent (cyan) and permanent (red) purchase hotspots in Istanbul.

For each customer it is known his/her income, age, and level of education, i.e. unknown (reported as 0), non-educated (as 1), elementary school (as 2), middle school (as 3), high school diploma (as 4), college (as 5), university degree (as 6), master degree (as 7), and Ph.D. (as 8). Moreover, it is known the location of customers' home ($home_u$), workplace ($work_u$), and transactions ($shop_{ui}$). Using such information it is possible to calculate the bin-wise distance of each purchase event (Eq. 4) as the minimum distance $d(home_u, shop_{ui})$ between the shop and the workplace, and the distance $d(work_u, shop_{ui})$ between the shop and the home (Singh, Bozkaya, & Pentland, 2015).

$$Dist_{ui}=\min(d(home_u, shop_{ui}), d(work_u, shop_{ui})) \quad (4)$$

By using the purchase distance, it is possible to evaluate few behavioral traits of the customers: (i) how *exploratory* each customer is when looking for places to shop, measured as the average purchase distance (*avgDist* in Fig. 6); and (ii) how *erratic* each customer is when looking for places to shop, measured as the standard deviation of the purchase distance (*stdDist* in Fig. 6).

Fig. 6 shows the results obtained from this analysis by means of violin plots. The violin plots represent the distribution of the demographic features for the population purchasing in each type of hotspot. As demographic features, we show individuals' education level (*edu* in Fig. 6), income, age, and the metrics for exploratory and erratic behavior. Overall their distribution is not significantly different, exception made for their average income, which is greater in the case of intermittent hotspots.

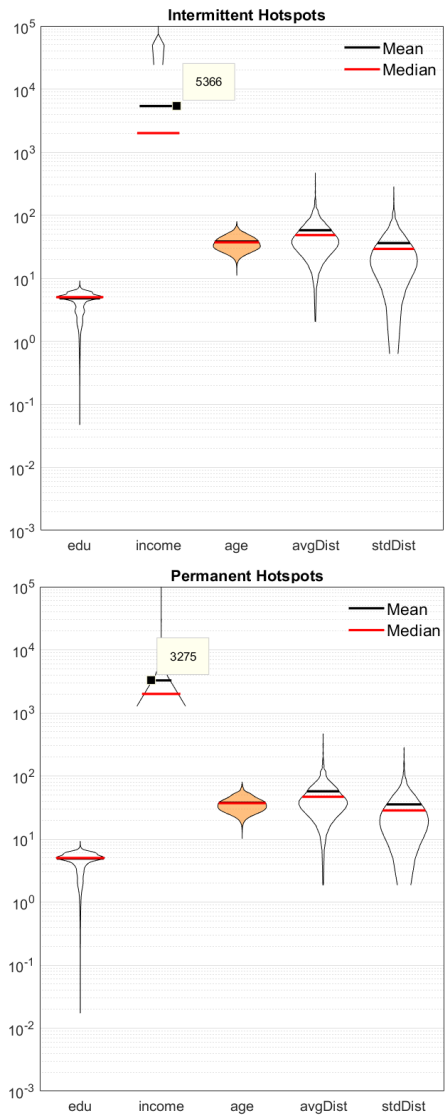

Fig. 6. Demographic features of the population purchasing in the hotspots. Log scale. The average income of the population purchasing in each type of hotspot is reported in squares.

The different average incomes among the people using the intermittent and permanent hotspots can be explained by considering the number of credit card transactions per customer, which is relatively low, i.e. on average 68 transactions per user throughout the year of observation. Therefore, for a hotspot to be "permanent" it must be able to attract as many people as possible, and this results in an average income closer to the one of the whole dataset, i.e. 2979 Turkish liras. On the other hand, intermittent hotspots are associated to occasional activities such as those related to nightlife, and these activities are more likely to be performed by people with higher incomes.

## 5 CONCLUSIONS

In this work, we have proposed an approach based on *Computational Stigmergy* to detect and analyze purchase hotspots according to their spatial distribution and temporal occurrence. The presented approach overcomes the limitations of many approaches in the literature that is the poor representation of the temporal dynamic and the inadequate exploration of the solutions space. By using our approach, we analyze the occurrence of the transactions of 60k Istanbul residents between 2014 and 2015. The results of such analysis confirm the validity of our approach to identify permanent and intermittent hotspots. Moreover, the analysis of the users spending in each type of hotspot is provided. As future works we aim at (i) studying the attractiveness of the areas identified as hotspots using a gravity model, and (ii) carrying out an analysis of the anomalies in the purchase activity, such as the one detectable by observing Fig. 4, in the group time slots immediately ahead of the 200th; apparently those are characterized by intermittent hotspots very similar to each other but completely different from all the others in any time slot, therefore a potential anomaly in the purchase activity.